\documentstyle[psfig,preprint,prl,aps]{revtex}
\def\r2{\sqrt 2}
\def\half{\frac{1}{2}}
\def\Acp{{\cal A}_{CP}}
\def\bsg{b \to s \gamma}
\def\bsgg{b \to s \gamma g}
\def\BSgamma{B \to X_s \gamma}
\def\sw2{\sin^2\theta_W}

\def\tb{\tan\beta}
\def\HC{H^\pm}
\def\w{\omega}

\def\m#1{{\widetilde m}_#1}
\def\mH{m_H}
\def\M{\widetilde M}
\def\mgr{m_{3/2}}

\def\PRD#1#2#3{Phys. Rev. {\bf D#1}, #3 (19#2)}
\def\NPB#1#2#3{Nucl. Phys. {\bf B#1}, #3 (19#2)}

\def\PLB#1#2#3{Phys. Lett. {\bf B#1}, #3 (19#2)}
\def\PRL#1#2#3{Phys. Rev. Lett. {\bf #1}, #3 (19#2)}
\def\PRep#1#2#3{Phys. Rep. {\bf #1}, #3 (19#2)}
\begin{document}
\draft
%\preprint{ICRR-Report-436-98-32} 
%\preprint{KEK-TH-604} 
%\preprint{OCHA-PP-128} 

\title{ 
Decay rate asymmetry in $\BSgamma$ as a signature of supersymmetry
}
\author{
Mayumi Aoki\cite{byline}
}
\address{
 Graduate School of Humanities and Sciences, Ochanomizu University  \\
Otsuka 2-1-1, Bunkyo-ku, Tokyo 112-8610, Japan  \\
}
\author{
Gi-Chol Cho\cite{byline}
}
\address{
 Theory Group, KEK, Tsukuba, Ibaraki 305-0801, Japan
}
\author{
Noriyuki Oshimo
}
\address{
Institute for Cosmic Ray Research, University of Tokyo  \\
Midori-cho 3-2-1, Tanashi, Tokyo 188-8502, Japan  \\
}
\date{\today}
\maketitle
\begin{abstract}
     In the supersymmetric standard model, 
the radiative $B$-meson decay $\BSgamma$ could receive a large 
contribution from a new source of flavor-changing neutral current,  
which also violates CP invariance.  
The CP-conjugate processes then may sizably differ in decay width.   
If a new CP-violating phase is not suppressed,  
this decay rate asymmetry can be significantly larger than the 
prediction by the standard model, possibly detectable at B factories.    
Such a large asymmetry may be implied by the measured branching ratio 
which is consistent with the standard model.  
\end{abstract}
\pacs{11.30.Er, 12.60.Jv, 13.25Hw}

% body of paper here
\narrowtext
     One of the main subjects at B factories is the measurement of
the unitarity triangle or the Cabibbo-Kobayashi-Maskawa (CKM) matrix.
Detailed analyses of the CKM matrix serve examinations of the
mechanisms for flavor-changing neutral current (FCNC) and CP violation
in the standard model (SM), through which the SM may be further confirmed,
or new physics may be revealed~\cite{nir}.
However, certain plausible extensions of the SM do not cause 
significant effects on the unitarity triangle.  
Although these non-standard effects may be extracted through various 
measurements of the sides and angles of the triangle,  
other studies should also be performed to search 
for new physics at B factories.

     The radiative $B$-meson decay $\BSgamma$ is a phenomenon 
which is sensitive to physics at the electroweak scale.  
Owing to a large mass of the $b$ quark, the inclusive decay $\BSgamma$ is 
well described by the free $b$-quark decays $\bsg$ and $\bsgg$, 
to which new physics could give sizable contributions through  
new interactions.  Then, the branching ratio of $\BSgamma$ 
may deviate from the SM prediction.  Furthermore, if CP 
invariance is violated by a new origin, 
which often occurs in extensions of the SM, a decay rate asymmetry 
\begin{equation}
\Acp = \frac{
   \Gamma(\overline{B}\to X_s\gamma)-\Gamma(B\to X_{\bar s}\gamma)}
   {\Gamma(\overline{B}\to X_s\gamma)+\Gamma(B\to X_{\bar s}\gamma)}  
\label{eq:Acp}
\end{equation}
may be generated sizably.  
Experimentally, the branching ratio has been measured to give \cite{cleo} 
\begin{equation}
{\rm Br}(\BSgamma) = (3.15 \pm 0.35 \pm 0.32 \pm 0.26) \times 10^{-4}, 
\end{equation}
and for $b$ hadrons produced at the $Z$ resonance \cite{aleph} 
\begin{equation}
{\rm Br}(H_b\to X_s\gamma) = (3.11 \pm 0.80 \pm 0.72) \times 10^{-4}. 
\end{equation}
These results are compatible with each other and 
do not show a large deviation from the SM prediction \cite{kagan1}.  
The decay rate asymmetry has not been measured.  Its predicted 
value in the SM is smaller than 0.01 \cite{soares}, while it may be 
enhanced in extensions of the SM~\cite{wolfenstein,kagan2,chua}.  

     In this letter, we discuss the decay rate asymmetry of $\BSgamma$
aiming at searching for supersymmetry.  
This radiative decay could receive nonnegligible contributions from 
new sources of FCNC contained in the supersymmetric standard 
model (SSM) \cite{misiak}.     
Under ordinary assumptions for the SSM, which we adopt, 
the one-loop diagram with up-type squarks and 
charginos and that with up-type quarks and charged Higgs bosons  
could sizably affect the decays $\bsg$ and $\bsgg$ \cite{oshimo}.  
Since the interactions of the charginos, up-type squarks, and 
down-type quarks induce also CP violation \cite{kizukuri,aoki} 
irrespectively of the standard KM mechanism, 
a large value of the decay rate asymmetry may be expected.  
We show that if a CP-violating phase in the squark mass-squared matrix  
is of order unity, the asymmetry can be much larger in the SSM than 
in the SM, providing a clear signature of supersymmetry.  
This large asymmetry is accommodated to the experimental results for   
the branching ratio, which may even suggest such a large value.  
The SSM could also give new contributions to $B^0$-$\bar B^0$ 
mixing, although their effects can only be observed indirectly 
through the values of the CKM matrix elements~\cite{branco}.  

     We assume that the masses and mixings of the squarks are described by 
the model based on $N=1$ supergravity and grand unification \cite{nilles}.  
At the electroweak scale, the squarks of  
interaction eigenstates are mixed in generation space.   
The generation mixings among the up-type squarks are approximately 
lifted by using the same matrices that diagonalize the mass 
matrix of the up-type quarks.  
As a result, the generation mixings in 
the interactions between down-type quarks and  
up-type squarks of mass eigenstates can be expressed 
by the CKM matrix $V$ of the quarks.  

     For the top squarks, the left-handed and right-handed 
ones are also mixed through the Yukawa couplings.  
The left-right mixings for the squarks of the first two generations 
can be neglected,  because of the smallness of the corresponding quark masses.  
The mass-squared matrix for the top squarks 
is given by   
\begin{eqnarray}
       M_t^2 &=& \left(
	\begin{array}{cc}
	\M_{uL}^2 + (1-c)m_t^2 & m_t(A^* \mgr + \cot\beta \mH) \\
	m_t(A \mgr + \cot\beta \mH^* ) & 
			\M_{uR}^2 + (1-2c) m_t^2 
	\end{array}
	\right), 
\label{eq:stopmass}
\end{eqnarray}
where $\mgr$ and $\mH$ denote the gravitino mass and the 
Higgsino mass parameter, respectively, and $\tb$ represents 
the ratio of the vacuum expectation values of the Higgs bosons.  
The dimensionless constant $A$ is related to the breaking of local 
supersymmetry above the grand unification scale and of order unity;  
$c$ is related to radiative corrections to the 
squark masses and has a value of $0.1-1$.  
The masses of the left-handed and right-handed up-type squarks 
for the first two generations are denoted by 
$\M_{uL}$ and $\M_{uR}$, which satisfy 
$\M_{uL}\simeq \M_{uR}\simeq |A|\mgr$.   
The down-type quarks and the up-type squarks couple to 
the charginos, whose mass matrix is given by 
\begin{eqnarray}
          M^- &=& \left( 
             	\begin{array}{cc}
               	\m2 & -\r2\cos\beta M_W \\
               	-\r2\sin\beta M_W & \mH
	              \end{array}
                          	\right),     
\label{eq:charginomass}
\end{eqnarray}
$\m2$ being the SU(2) gaugino mass.  

     The new sources of CP violation reside in   
the top-squark mass-squared matrix $M_t^2$ and the chargino 
mass matrix $M^-$.  For physical complex parameters,     
without loss of generality, we can take the dimensionless constant $A$ 
and the Higgsino mass parameter $\mH$, which are expressed as    
\begin{equation}
A = |A| \exp(i\alpha), \quad \mH = |\mH| \exp(i\theta).  
\end{equation} 
 However, the CP-violating phase $\theta$ is severely 
constrained by the experimental bounds on the electric 
dipole moments (EDMs) of the neutron and the electron \cite{kizukuri}.  
If $\theta$ is of order unity, 
the squarks and sleptons are not allowed to have masses smaller than 1 TeV.  
Then, the decay $\BSgamma$ receives only small contributions from 
the chargino-squark loop diagram, leading to negligible CP violation 
in this process.  
On the other hand, for a sufficiently small magnitude of $\theta$ 
with $\m2\gtrsim 500$ GeV, the squarks 
and sleptons can be of order 100 GeV  while another CP-violating 
phase $\alpha$ being unsuppressed \cite{aoki}.  
In this parameter region the SSM 
may induce sizable CP violation without causing discrepancies for the EDMs.     
  
     For the decays $\bsg$ and $\bsgg$, the effective Hamiltonian 
with five quarks, the heavier degrees of freedom being integrated out, 
is written generally by \cite{grinstein} 
\begin{eqnarray}
H_{eff} &=& -\frac{4G_F}{\r2} V_{ts}^*V_{tb}\sum_{j=1}^8 C_j(\mu) O_j(\mu),  \\
O_2 &=& \overline{s_L}\gamma_\mu c_L\overline{c_L}\gamma^\mu b_L, 
                        \nonumber    \\
O_7 &=& \frac{e}{16\pi^2}m_b\overline{s_L}\sigma^{\mu\nu}b_R F_{\mu\nu},  
                        \nonumber  \\
O_8 &=& \frac{g_s}{16\pi^2}m_b\overline{s_L}\sigma^{\mu\nu}T^ab_R 
             G_{\mu\nu}^a,  \nonumber  
\end{eqnarray}
where $C_j(\mu)$ denotes a Wilson coefficient evaluated at the scale $\mu$, 
and $F_{\mu\nu}$ and $G_{\mu\nu}$ respectively represent the electromagnetic 
and strong field strength tensors, $T^a$ being the generator of SU(3).   
The contributions from the operators other than $O_2$, $O_7$, and $O_8$  
are negligible.   

     At the electroweak scale, the Wilson coefficients $C_7$ and $C_8$ 
receive contributions at the one-loop level.  Sizable new contributions 
by the SSM come from the diagrams in which the charginos $\w$ or the charged 
Higgs bosons $\HC$ are exchanged together with the up-type squarks 
or the up-type quarks, respectively.   
We express the coefficients as the sum of dominant contributions  
\begin{equation}
C_j(M_W) = C_j^W(M_W)+C_j^{\HC}(M_W)+C_j^\w(M_W),  
\end{equation}
where $C_j^W$ stands for the standard $W$-boson contribution.  
The leading-order (LO) calculations give 
\begin{eqnarray}
C_7^W(M_W) &=& -\frac{3}{2} r_W \bar K_1(r_W),  \nonumber \\
C_7^{\HC}(M_W) &=& -\half r_H \biggl\{ \cot^2\beta 
         \bar K_1(r_H) + \bar K_2(r_H)\biggr\},  \nonumber \\ 
C_7^\w(M_W) &=& \sum_{i=1}^2 \frac{M_W^2}{m_{\w i}^2}
	\biggl[ -|C_{R1i}|^2 r_{ui} K_1(r_{ui}) 
	- \frac{C_{R1i} C_{L2i}^*}{\r2 \cos\beta} 
	\frac{m_{\w i}}{M_W}r_{ui} K_2(r_{ui})              \nonumber \\
	&&	+ \sum_{k=1}^2 \biggl\{ \biggl| C_{R1i} S_{t1k} 
	- \frac{C_{R2i}S_{t2k}}{\r2 \sin\beta} \frac{m_t}{M_W} \biggr|^2
	r_{ki} K_1(r_{ki})            \nonumber \\
	&&+ \frac{ C_{L2i}^* S_{t1k}^* }{\r2 \cos\beta} 
	\biggl(C_{R1i} S_{t1k} - \frac{C_{R2i}S_{t2k}}{\r2 \sin\beta}
	\frac{m_t}{M_W}\biggr) \frac{m_{\w i}}{M_W} r_{ki} 
	K_2(r_{ki}) \biggr\} \biggr], \nonumber \\ 
  \bar K_a(r) &=& \frac{2}{3}I_a(r)+J_a(r), \quad 
  K_a(r) = I_a(r)+\frac{2}{3}J_a(r),   
\label{eq:formfactor}   \\
  r_W &=& \frac{m_t^2}{M_W^2}, \quad r_H=\frac{m_t^2}{M_{\HC}^2}, 
 \quad r_{ui}=\frac{m_{\w i}^2}{\M_{uL}^2}, 
 \quad r_{ki}=\frac{m_{\w i}^2}{\M_{tk}^2}, 
                           \nonumber  
\end{eqnarray}
where $S_t$ and $C_L$, $C_R$ are the unitary matrices which 
diagonalize $M_t^2$ and $M^-$, respectively.  The functions $I_a$ 
and $J_a$ are defined in Ref.~\cite{oshimo}.    
The LO contributions $C_8^W(M_W)$, $C_8^{\HC}(M_W)$, 
and $C_8^\w(M_W)$ are obtained by replacing $\bar K_a$ and $K_a$ 
by $I_a$ and $J_a$, respectively, in Eq.~(\ref{eq:formfactor}).  
In our scheme for the SSM, the one-loop diagram mediated by the gluinos or the 
neutralinos with the down-type squarks cause only small effects 
on both FCNC and CP violation~\cite{oshimo,kizukuri}.  

     The contributions $C_7^\w(M_W)$ and $C_8^\w(M_W)$ have complex values, 
which are traced back to physical complex phases intrinsic in the SSM.  
Consequently, CP invariance is violated in the decay $\BSgamma$ 
irrespectively of the KM mechanism.   
Another important effect by the SSM is that 
$C_7^\w(M_W)$ and $C_8^\w(M_W)$ can be added to $C_7^W(M_W)$ and 
$C_8^W(M_W)$ both constructively and destructively depending on 
the parameter values \cite{oshimo}, 
whereas $C_7^{\HC}(M_W)$ and $C_8^{\HC}(M_W)$ are added constructively.  
These effects make it possible to have a large magnitude for 
the decay rate asymmetry while keeping the branching ratio 
comparable with the SM value.  
     
     Given the Wilson coefficients at $\mu=M_W$, those at 
$\mu=m_b$ are obtained by solving the renormalization-group equations.  
At the LO,  these coefficients are given by   
\begin{eqnarray}
      C_2(m_b) &=& \half (\eta^{-\frac{12}{23}}+\eta^{\frac{6}{23}}), 
                     \nonumber \\
      C_7(m_b) &=& \eta^{\frac{16}{23}}C_7(M_W) + 
	 \frac{8}{3} (\eta^{\frac{14}{23}} - \eta^{\frac{16}{23}}) C_8(M_W) 
          + \sum_{i=1}^8 h_i\eta^{a_i},  \nonumber \\
     C_8(m_b) &=& \eta^{\frac{14}{23}}C_8(M_W) 
      + \sum_{i=1}^8 \bar h_i\eta^{a_i},  
\end{eqnarray}
with $\eta=\alpha_s(M_W)/\alpha_s(m_b)\approx 0.56$,   
where $h_i$, $\bar h_i$, and $a_i$ are numerical constants 
defined in Ref.~\cite{buras}.   

     The decay rate asymmetries of $\bsg$ and $\bsgg$ are 
induced by the interferences between 
the tree level processes and the one-loop level processes with 
absorptive parts.  
Combining these asymmetries, the asymmetry for $\BSgamma$ is 
given by \cite{kagan2,greub}  
\begin{eqnarray}
     \Acp &=& \frac{4\alpha_s(m_b)}{9|C_7(m_b)|^2}
   \biggl(\biggl[ \frac{10}{9}-2z\{v(z)+b(z,\delta)\}\biggr]
                     {\rm Im}\biggl[C_2(m_b)C_7^*(m_b)\biggr]   \nonumber \\
   & & - {\rm Im}\biggl[C_8(m_b)C_7^*(m_b)\biggr] 
     + \frac{2}{3}zb(z,\delta){\rm Im}\biggl[C_2(m_b)C_8^*(m_b)\biggr]\biggr),  
\label{eq:cpasy2}
\end{eqnarray}
with $z=m_c^2/m_b^2\approx (0.29)^2$, 
where we have neglected the small effects arising from CP violation 
by the standard KM mechanism.  
The functions $v(z)$ and $b(z,\delta)$ are defined in Ref. \cite{kagan2}.  
The inclusive decay $\BSgamma$ includes the three-body decay $\bsgg$, 
which leads to a continuous energy spectrum for the photon.  
The asymmetry in Eq.~(\ref{eq:cpasy2}) is calculated with the 
photon energy being cut as $E_\gamma>(1-\delta)m_b/2$.  

     The parameter values of the SSM are constrained by the measured 
branching ratio of $\BSgamma$ as well as direct searches for 
supersymmetric particles.  The full next-to-leading order (NLO) calculations 
for this decay width have already been performed in the SM.   
However, for the SSM, the NLO matching conditions of 
$C_7^\w$ and $C_8^\w$ at $\mu=M_W$ have not yet obtained in general form.  
We thus calculate the decay width of $\BSgamma$ by  
using the matrix elements and anomalous dimensions at the NLO, and also 
the matching conditions for $C_7^W$ and $C_8^W$ at the NLO, while 
those for the chargino and charged Higgs boson contributions at the LO.  
The branching ratio is obtained by the usual procedure of 
normalizing the decay width to that of the semileptonic decay 
$B\to X_ce\nu$, whose branching ratio is taken for 0.105.   

     We show in Fig. \ref{fig:asymmetry} the decay rate 
asymmetry (upper) and the branching ratio (lower) 
as a function of the lighter top squark mass $\M_{t1}$, 
with $c=0.1-1$ and $\M_{uL}\simeq\M_{uR}\simeq|A|\mgr$.  
The range $\M_{t1}\lesssim 80$ GeV is ruled out by LEP2 
experiments \cite{lep}.  
The CP-violating phases are set for $\alpha=\pi/4$ and $\theta=0$.  
The other parameter values are taken for 
$\m2=500$ GeV, $\mH=100$ GeV, $M_{\HC}=200$ GeV, and $\tb=10$.  
The energy-cut parameter is set for $\delta=0.99$.  
In the lower figure, the experimental bounds of the branching 
ratio~\cite{cleo,aleph} are also indicated.  
For 100~GeV~$\lesssim \M_{t1}\lesssim$~400~GeV, the branching ratio lies within 
the experimental bounds by ALEPH,
where the asymmetry has a value $-0.07\lesssim\Acp\lesssim-0.02$.  
As the charged Higgs boson mass $M_{\HC}$ decreases, the branching 
ratio increases, and the allowed range for $\M_{t1}$ becomes narrower.   
The asymmetry does not vary much with $M_{\HC}$.     
The peaks of the asymmetry and the branching ratio are both 
roughly at the same value of $\M_{t1}$, which increases with $\tb$ 
for $1<\tb<30$.  The values of the peaks do not depend 
significantly on $\tb$. (Detailed numerical analyses will be presented 
elsewhere~\cite{aoki2}.)      

     In general, as long as the CP-violating phase $\alpha$ is not suppressed 
and the charged Higgs boson mass is of order 100 GeV, 
the magnitude of the asymmetry is larger than 0.01 in the parameter region 
where the branching ratio is consistent with the experiments.     
For such a nonheavy mass of $\HC$, the sum of the contributions 
of $W$ and $\HC$ alone makes the decay width too large \cite{borzumati}.  
Therefore, in the SSM parameter region allowed by the branching ratio, 
the chargino contribution has to be comparable with the contributions 
of $W$ and $\HC$, leading to a large CP asymmetry.   
If the charged Higgs boson is sufficiently heavy, its contribution 
to the decay width is negligible.  Still, there are parameter regions 
where the asymmetry is larger than 0.01 without conflicting with 
the measured branching ratio.   
It was recently claimed \cite{chua} that the magnitude of $\Acp$ can 
significantly exceed 0.1 by the contribution of the gluino-squark 
loop diagram.  
However, this contribution is much smaller than that from 
the chargino-squark loop diagram in our scheme.  

     At B factories, $B\bar B$ pairs are expected to be produced 
to the amount of order $10^8$ per year \cite{nir}, 
among which approximately half are pairs of $B^+B^-$.  
Tagging $B^+$ or $B^-$ by the semileptonic decays 
$B\to X_ce\nu_e, X_c\mu\nu_\mu$,  
whose branching ratio is about 0.1 each, 
roughly $2\times 10^{3}$ events of $\BSgamma$ would be observed  
with the charge of $B$ being known.  
Then, the decay rate asymmetry with $|\Acp|> 0.2$ could be measurable 
in experiments.    
The charge of $B$ may be also identified by $X_s$ itself, 
making it possible to improve the measurability of the asymmetry.   
There is a sizable region of parameter space in the SSM 
where the asymmetry can be detected.  

     In summary, we have discussed the decay rate asymmetry for 
the radiative $B$-meson decay $\BSgamma$ in the SSM based on $N=1$ 
supergravity.  
Among the possible new contributions in this model, the chargino 
and the charged Higgs boson loop diagrams yield sizable effects.  
Assuming an unsuppressed CP-violating phase intrinsic in the SSM, 
the asymmetry can have a large value, maximally of order 0.1, 
not expected by the SM.  
In particular, the experimental results that the branching ratio is 
consistent with the SM would imply  
a large asymmetry.   
Measuring the decay rate asymmetry of $\BSgamma$ may 
provide a first hint for supersymmetry.  

%\acknowledgements
\smallskip

     The authors acknowledge discussions with 
J. Arafune, T. Goto, and Y. Shimizu.  
The works of M.A. and G.-C.C. are supported in part by 
the Grant-in-Aid for Scientific 
Research from the Ministry of Education, Science and Culture, Japan.  

% now the references. delete or change fake bibitem. delete next three
%   lines and directly read in your .bbl file if you use bibtex.
%

% Put the label that you will use with \ref{} command in 
% the braces of the \label{} command.
%
% \begin{figure}
% \caption{}
% \label{}
% \end{figure}
%
\begin{figure}
\caption{The decay rate asymmetry (upper) and the branching ratio (lower)
 for $\BSgamma$. }
\label{fig:asymmetry}
\end{figure}
\begin{figure}
\begin{center}
\leavevmode
\psfig{file=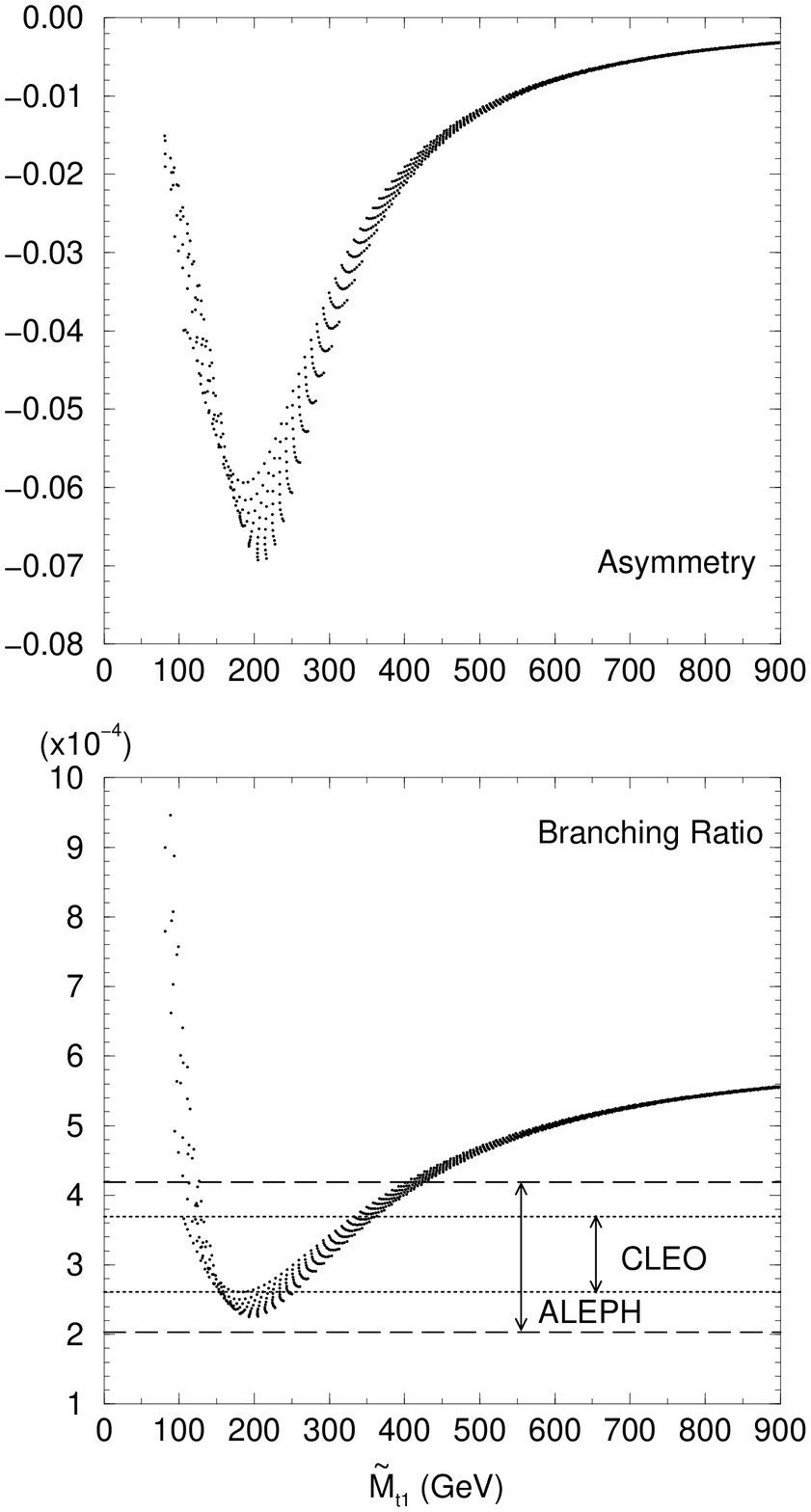,height=17cm}
%%%  options --> height, width, angle
%%%
\end{center}
\end{figure}
\pagebreak
% Put the label that you will use with \ref{} command in the 
% braces of the \label{} command.
% Insert the column specifiers (l, r, c, d, etc.) in 
% the empty braces of the \begin{tabular}{} command.
%
% \begin{table}
% \caption{}
% \label{}
% \end{tabular}
% \end{table}
%   
\end{document}